\documentclass[a4paper,12pt,preprint,amsmath,showpacs,
nofootinbib,superscriptaddress]{revtex4}
\usepackage[hypertex]{hyperref}
\usepackage{graphicx}
\usepackage{epstopdf}
\usepackage{bm}
\usepackage{epsfig}
\usepackage{graphics}
\usepackage{xspace}
\usepackage{amsfonts}

\def\Slash#1{{#1\!\!\!\slash}}

\newcommand{\nn}{\nonumber}

\newcommand{\ep}{\epsilon}

\newcommand{\bea}{\begin{eqnarray}}
\newcommand{\eea}{\end{eqnarray}}

\begin{document}


\title{ Next-to-leading order QCD predictions for the signal of Dark Matter and
 photon associated production at the LHC }

\vspace*{1cm}

\author{Jian Wang}
\affiliation{Department of Physics and State Key
Laboratory of Nuclear Physics and Technology, Peking
University, Beijing, 100871, China}

\author{Chong Sheng Li\footnote{Electronic
address: csli@pku.edu.cn}}
\affiliation{Department of Physics and State Key
Laboratory of Nuclear Physics and Technology, Peking
University, Beijing, 100871, China}
\affiliation{Center for High Energy Physics, Peking
University, Beijing, 100871, China}

\author{Ding Yu Shao}
\affiliation{Department of Physics and State Key
Laboratory of Nuclear Physics and Technology, Peking
University, Beijing, 100871, China}

\author{Hao Zhang}
\affiliation{Department of Physics and State Key
Laboratory of Nuclear Physics and Technology, Peking
University, Beijing, 100871, China}



\begin{abstract}
 \vspace*{0.3cm}
We study the potential of the LHC to discover the signal of Dark Matter
associated production with a photon induced by a dimension six
effective operator, including NLO QCD corrections. We investigate the
main backgrounds from SM, i.e. $Z$ boson and a photon
associated production with invisible decay of $Z$ boson, and $Z$ boson
and a jet production with the jet misidentified as a photon. We find
that the $p_T^{\gamma}$ distributions of the backgrounds decrease
faster than that of the signal with increasing of the transverse
momentum of the photon. The $\eta^{\gamma}$ distributions of the
backgrounds are almost flat in the full range of $\eta^{\gamma}$.
In contrast, the signal lies mainly in the central region of
$\eta^{\gamma}$. These characteristics may help to select the events
in experiments. We show that in the parameter space allowed by the
relic abundance constraint, which we have calculated at the  NLO QCD
level, the LHC  with $\sqrt{S}=7{\rm ~TeV}$ may discover this
signal at the $5\sigma$ level after collecting an integrated
luminosity of $1~{\rm fb}^{-1}$. On the other hand, if this signal is not
observed at the LHC, we can set a lower limit on the new physics scale at the $3\sigma$ level.

\end{abstract}

\pacs{12.38.Bx, 95.35.+d, 14.70.Bh, 14.65.Jk}

\maketitle
\newpage


\section{Introduction}
\label{sec:1}

Astrophysical and cosmological observations have confirmed the
existence of Dark Matter (DM) in our universe\cite{Jungman:1995df}
and the density of DM is much larger than that of
the visible matter\cite{Komatsu:2008hk}. The relic abundance of DM favours a
weakly interacting massive particles (WIMP). This kind of DM has
been extensively studied in the literatures. Since there are no
candidates of this kind of DM in the Standard Model (SM), any
discovery of the signal of DM imply new physics.

In the region of DM accumulating, DM can annihilate into SM particles,
such as photons, electrons and positrons. These produced
particles can propagate through the interstellar space and be
detected by experiments on the earth, such as
PAMELA\cite{Adriani:2010rc}, ATIC\cite{:2008zzr},
HESS\cite{Aharonian:2008aa} and Fermi LAT\cite{Abdo:2009zk}. The
detection of these signals is not conclusive evidence for DM, since
it depends on the assumptions of the distribution of DM and the
propagator model. Besides, other astrophysical interpretation can
not be excluded. Another way to look for the signal of DM is to
measure the recoil energy of nuclei caused by the elastic scattering
of a WIMP off a nucleon, such as the experiments of
DAMA\cite{Bernabei:2008yi}, CDMS\cite{Ahmed:2008eu},
CoGeNT\cite{Aalseth:2010vx} and XENON\cite{Aprile:2010um}. The DAMA
and CoGeNT experiments favour a light DM with a mass around $10$~GeV.
The CDMS and XENON experiments set upper limits on the WIMP
and nucleon spin-dependent and spin-independent cross sections if
the mass of the WIMP ranges from $10$~GeV to $1000$~GeV. These
experiments are passive and much time was spent in waiting for the
collision with  the DM. More active approach is to produce the DM in
the laboratory directly, such as the Large Hadron Collider (LHC) if
DM exists and has interactions with the SM particles. There are a
lot of studies to search for DM at the LHC
in varies of DM models\cite{Buchmueller:2011ki,Li:2011in,
Profumo:2011zj,Belanger:2011ny,Kile:2011mn,Akula:2011dd,Feldman:2011me,
Bai:2010hd,Gogoladze:2010ch,Cheung:2010zf,Goodman:2010ku,Bertone:2010rv,
Giudice:2010wb,Li:2010rb,Beltran:2010ww,Zhang:2009dd,ArkaniHamed:2008qp,
Fargion:1995qb}.

Because the LHC is a proton-proton collider, the QCD correction should be
considered for any process if people want to make a reliable prediction.
In this work, using model independent method, we investigate the
possibility of discovering the DM in associated production with a
photon induced by a dimension six effective operator at the
next-to-leading (NLO) order QCD level, since this signal is clear
and suffer from little backgrounds from the SM.

This paper is organized as follows. In section~\ref{sec:operator},
we describe the dimension six effective operator for this process.
In section~\ref{sec:relic}, we calculate the relic abundance induced
by this effective operator and find the allowed region for the mass
and couplings of the DM. In section~\ref{sec:nlo}, we present the
details of the NLO QCD corrections to the associated production of
the DM and photon and discuss the dependence of the K-factor on the
mass and couplings of the DM. In section~\ref{sec:background}, we
calculate the backgrounds in SM and analyze the discovery potential
at the LHC. Conclusion will be given in
section~\ref{sec:conclusion}.

\section{Effective Operator}
\label{sec:operator}
The DM studied in this work is a Dirac fermion, denoted by $\chi$. It is a singlet under
the SM gauge group $SU(3)_{c}\times SU(2)_{L} \times U(1)_{Y}$. As a result, this
kind of DM does not participate in any strong or electroweak interactions. In
addition, we assume it couples with the SM quarks in the form
\begin{equation}\label{eq-operator}
   \mathcal{O}=\frac{\kappa}{\Lambda^{2}}(\bar{q}q)(\bar{\chi}\chi),
\end{equation}
where $\Lambda$ is the new physics scale above which new particles should appear.
This four-fermion operator has been
considered in Refs.\cite{Beltran:2008xg, Cao:2009uw, Goodman:2010ku, Bai:2010hh} in discussing
DM interactions. However, we will
focus on the signal of DM and photon associated production at the LHC which
has not been carefully investigated before. Moreover, we calculated the NLO QCD corrections
for this process whose effects are important for research at the LHC. Note that
the new physics scale $\Lambda$ can be viewed as the remnant of integrating
the propagator between the SM particles and DM. Therefore, this operator is valid
only if $\Lambda > m, \sqrt{\hat{s}}$, where $m$ is the mass of DM and $\sqrt{\hat{s}}$ the center
of mass energy of the collision.     
Generally, it is possible that $\sqrt{\hat{s}}>\Lambda$ at the LHC. 
However, the luminosity for the process drops very fast with
the increasing of $\sqrt{\hat{s}}$ at the LHC with
$\sqrt{S}=14~(7)~\rm TeV$, which can be seen in Fig.1.
Thus, we relax this limit in practical numerical calculation,
where we set the default value of $\Lambda$=500 GeV.

\begin{figure}
  \includegraphics[width=0.45\linewidth]{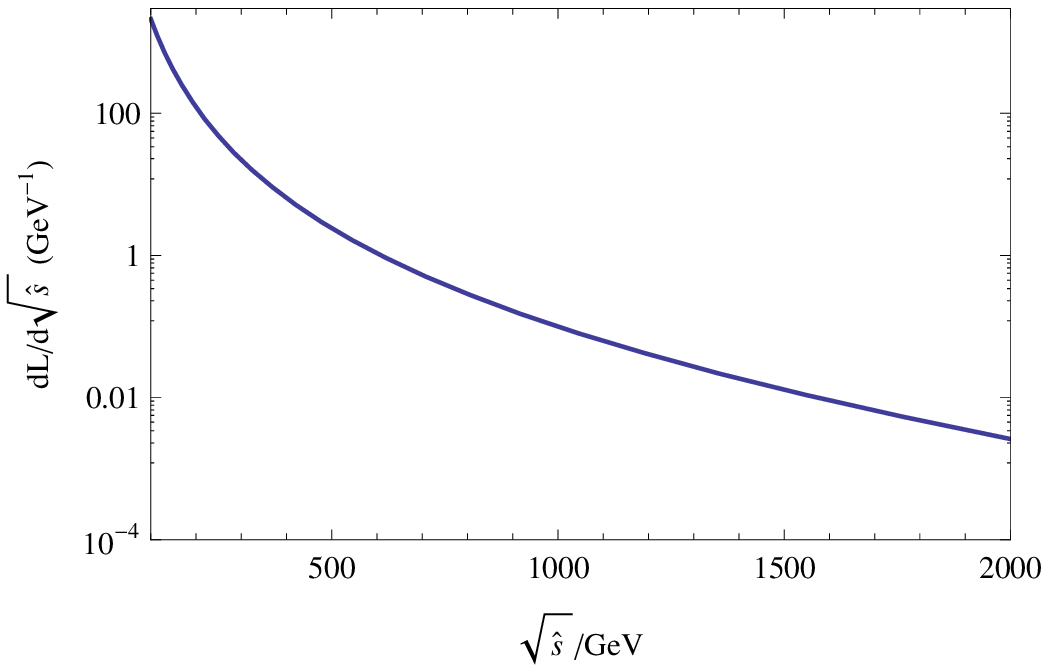}\includegraphics[width=0.45\linewidth]{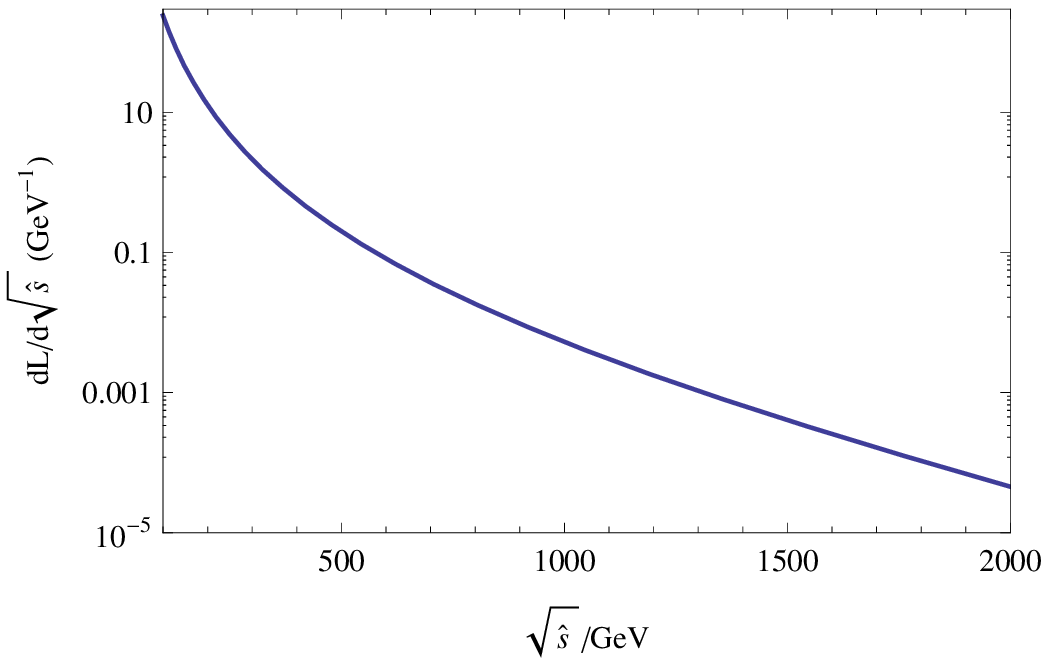}\\
  \caption{Luminosity plots for $u\bar{u}$ initial states at the LHC with $\sqrt{S}=14$ (left) and 7 (right) TeV.}
  \label{fig-lum}
\end{figure}

\section{Relic Abundance}
\label{sec:relic}
The DM relic abundance is a precision observable in cosmology.
The DM we choose to study can contribute to the relic abundance of cold Dark Matter (CDM).
Combining WMAP data with the latest distance
measurements from Baryon Acoustic Oscillations in the
distribution of galaxies and Hubble constant
measurements gives the constrain\cite{Jarosik:2010iu}
\begin{equation}\label{eq-relic}
    \Omega_{CDM}h^2=0.1123\pm 0.0035,
\end{equation}
where $\Omega_{CDM}$ is the CDM energy density of the Universe normalized by the critical density and
$h=0.710\pm 0.025$ is the scaled Hubble parameter.

The relic abundance can be calculated from the total annihilation cross section of DM.
First, we give the LO total annihilation cross section as
\begin{eqnarray}
  \sigma_B^{an} v =N_c N_f \frac{\kappa^2}{\Lambda^4}\frac{s-4m^2}{8\pi},
\end{eqnarray}
where $v$ is the relative velocity between the DM. $N_c$ and $N_f$ are the numbers of  color and
flavor of quarks, respectively. This result agrees with that in Ref.\cite{Beltran:2008xg}.

The NLO corrections to the total annihilation cross section comprise of two parts:
the one-loop virtual corrections and real gluon emission corrections. The results of
virtual corrections are
\begin{equation}
    \sigma_v^{an}=\sigma_B^{an,\epsilon}\left(\frac{\alpha_s C_F}{2\pi}\right)D_{\epsilon}
    \left[-\frac{2}{\epsilon_{IR}^2}-\frac{3}{\epsilon_{IR}}-3\ln\left(\frac{s}{\Lambda^2}\right)-2+\pi^2
    \right],
\end{equation}
where $D_{\epsilon}=(4\pi\mu^2/s)^{\epsilon}/\Gamma(1-\epsilon)$ and $\sigma_B^{an,\epsilon}$ is the n-dimensional $(n=4-2\epsilon)$
LO total annihilation cross section
\begin{equation}
    \sigma_B^{an,\epsilon}=\left(\frac{4\pi}{s}\right)^{\ep}\frac{\Gamma(1-\ep)}{\Gamma(2-2\ep)}\sigma_B^{an}.
\end{equation}
In obtaining the above results, we have renormalized the effective operator in the $\overline{\rm MS}$ scheme.
The results of real corrections are
\begin{equation}
    \sigma_r^{an}=\sigma_B^{an,\epsilon}\left(\frac{\alpha_s C_F}{2\pi}\right)D_{\epsilon}
    \left(\frac{2}{\epsilon_{IR}^2}+\frac{3}{\epsilon_{IR}}+\frac{21}{2}-\pi^2\right).
\end{equation}
Combining the two parts we get the NLO total annihilation cross section
\begin{equation}
    \sigma^{an}_{NLO}=\sigma_B^{an}\left[1+\frac{\alpha_s C_F}{2\pi}\left(\frac{17}{2}-3\ln\left(\frac{s}{\Lambda^2}\right)\right)\right].
\end{equation}

The DM  is moving at nonrelativistic velocities($v \ll 1$) when freezing out. Thus we can expand
\begin{equation}
  \sigma^{an} v = a + b v^2,
\end{equation}
where
\begin{eqnarray}
  a &=& 0, \nn \\
  b &=& K^{an}N_c N_f\frac{\kappa^2}{\Lambda^4}\frac{m^2}{8\pi},
\end{eqnarray}
in which $K^{an}$ is the K-factor of the DM annihilation cross section
\begin{equation}
    K^{an}=1+\frac{\alpha_s C_F}{2\pi}\left[\frac{17}{2}-3\ln\left(\frac{4m^2}{\Lambda^2}\right)\right].
\end{equation}

\begin{figure}
  \includegraphics[width=0.6\linewidth]{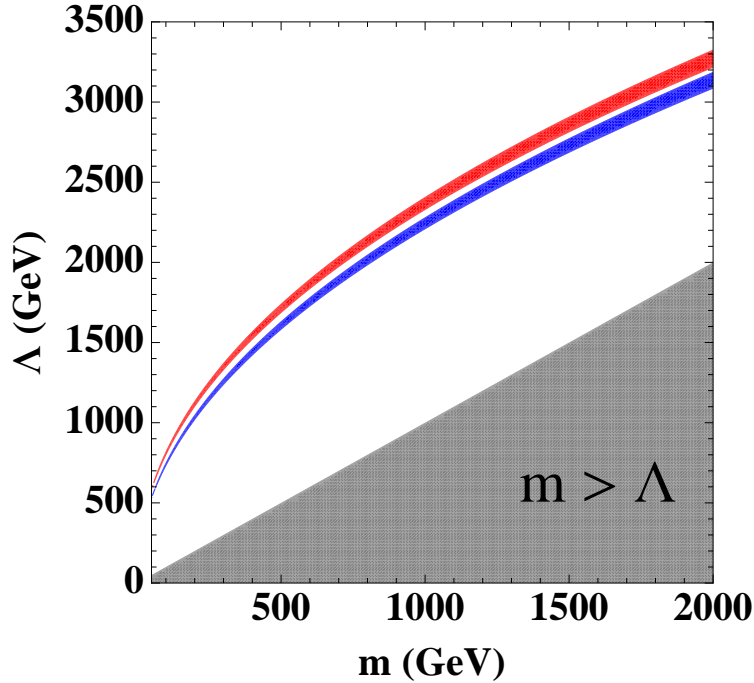}\\
  \caption{ Relation between the mass of DM and the
new physics scale. The relic abundance is required be to in the $2\sigma$ region
around the observed central value. The blue band is the LO result. The red band is
the NLO result. In this figure, we choose $\kappa=1, \alpha_s=0.118$ and $ N_f=5$. }
  \label{fig-relic}
\end{figure}

The freezeout epoch $x_f=m/T_f$ is determined by\cite{Kolb:1990vq}
\begin{equation}
  x_f = \ln\left[0.456 b m_{Pl}m\left(g/g_{\ast}^{1/2}\right)\right]-\frac{3}{2}\ln\left\{\ln\left[0.456 b m_{Pl}m\left(g/g_{\ast}^{1/2}\right)\right]\right\},
\end{equation}
where $m_{Pl}=1.22\times 10^{19} {\rm ~GeV}$ is the Planck mass. $g$ and $g_{\ast}$ is the number of relativistic degrees of freedom and
effective number of relativistic degrees of freedom at the freeze-out temperature $T_f$, respectively.
The result for the relic abundance is
\begin{equation}
    \Omega_{\chi}h^2\approx \frac{1.07\times 10^{9}{\rm ~GeV}^{-1}x_f^2}
  {3(g_{\ast S}/g_{\ast}^{1/2})m_{Pl}b}.
\end{equation}
For most of the history of the universe all particle species had a common temperature,
and $g_{\ast S}$ can be replaced by $g_\ast$.
Requiring the DM relic abundance is in the $2\sigma$ region
around the observed central value, the new physics scale of the effective operator is
determined by the mass of the DM. We show this relation in Fig.\ref{fig-relic}.
When the parameters $(m,~\Lambda)$ satisfy this relation, the NLO QCD K-factor is
nearly 1.4.
Comparing the LO result (blue band) and the NLO result (red band),
we find that the NLO QCD correction increases the new physics scale by about 10\%.
Since we do not hold the point of view that the abundance of the WIMP is determined only by
this one kind of DM, the regions below the red band are all allowed.

\section{NLO QCD Corrections to DM and photon associated production}
\label{sec:nlo}
According to the operator in (\ref{eq-operator}), the DM can be pair produced at
hadron colliders. However, because DM can not decay into SM particles, such processes
give just missing energy and no observable signals. As a result, we have to
consider the process of DM associated production with a photon or a
jet\cite{Cao:2009uw, Goodman:2010ku, Bai:2010hh}. Though the cross section of
 associated production with a photon is less than that of associated production with a jet,
the signal of this process is characteristic and bring with less backgrounds.

\subsection{LO calculation}
\label{subsec:21}
\begin{figure}
  \includegraphics[width=0.6\linewidth]{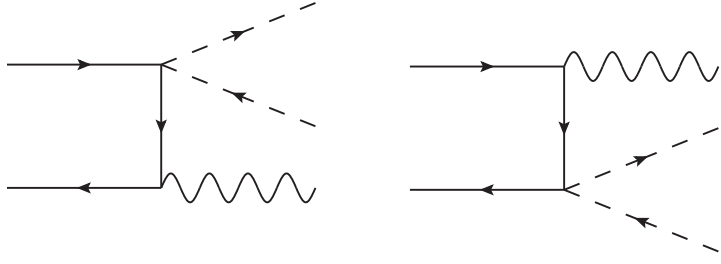}\\
  \caption{LO Feynman Diagrams.}
  \label{fig-born}
\end{figure}

We start from the leading order (LO) calculation. The LO Feynman diagrams for
this process
\begin{equation}\label{eq-process}
    q(p_1)+\bar{q}(p_2) \to \chi(p_3)+\bar{\chi}(p_4)+\gamma(p_5)
\end{equation}
are shown in Fig.\ref{fig-born}.

Because the DM does not interact with SM particles except for the
quark fields in the operator (\ref{eq-operator}),
their contributions to the cross section can be factorized as
\begin{equation}\label{eq-DM}
    |\mathcal{M}_{DM}|^2=2 \left(s_{34}-4 m^2\right).
\end{equation}
where $s_{ij}=(p_i+p_j)^2$.
Therefore, the born matrix element can be written as
\begin{equation}\label{eq-Mborn}
    \mathcal{M}_B=\frac{-2 M_3 t_{15}+M_1 \left(-t_{15}-t_{25}\right)+2 M_2 t_{25}}{t_{15} t_{25}}\mathcal{M}_{DM},
\end{equation}
where the $M_{i},i=1,2,3$ represent the standard matrix elements in our calculation which are defined as
\begin{eqnarray}\label{eq-stdmatrix}
    M_1 &=& \bar{v}(p_2)\Slash{p}_5\gamma^{\mu}u(p_1)\epsilon_{\mu}(p_5), \nn\\
    M_2 &=& \bar{v}(p_2)u(p_1)p_1^{\mu}\epsilon_{\mu}(p_5), \nn\\
    M_3 &=& \bar{v}(p_2)u(p_1)p_2^{\mu}\epsilon_{\mu}(p_5).
\end{eqnarray}

The spin and color summed and averaged Born matrix element squared  is
\begin{equation}\label{eq-Born}
    \overline{|\mathcal{M}_B|^2}=\frac{4\pi\alpha \kappa^2}{3\Lambda^4}\frac{s_{12}^2+s_{34}^2}
    { t_{15} t_{25}}|\mathcal{M}_{DM}|^2,
\end{equation}
where $t_{ij}=(p_i-p_j)^2$ and $\alpha=e^2/4\pi$. Then the LO partonic cross section is
\begin{equation}\label{eq-LOpartonic}
    \hat{\sigma}_B=\frac{1}{2s_{12}}\int d \Gamma_3 \overline{|\mathcal{M}_B|^2},
\end{equation}
in which $\Gamma_3$ is the three particle final states phase space. After convoluting with
the parton distribution functions (PDFs) $G_{q(\bar{q})}(x)$, we obtain the LO cross section
\begin{equation}\label{eq-LOhadronic}
    \sigma_B=\int d x_1 d x_2 [G_{q/p}(x_1)G_{\bar{q}/p}(x_2)+(x_1 \leftrightarrow x_2)]\hat{\sigma}_B.
\end{equation}
.

\subsection{ NLO results }
The NLO QCD corrections involve the one-loop virtual gluon
effects and contributions of real gluon
and quark or antiquark emissions  to leading order processes.
To deal with ultraviolet (UV) and infrared (IR) (soft and collinear) divergences in our computation,
we use $n=4-2\epsilon$ dimensional regularization to regulate these
divergences, and all divergences appear as $1/\epsilon^{i}$ with
$i=1,2$.

\begin{figure}
  \includegraphics[width=0.6\linewidth]{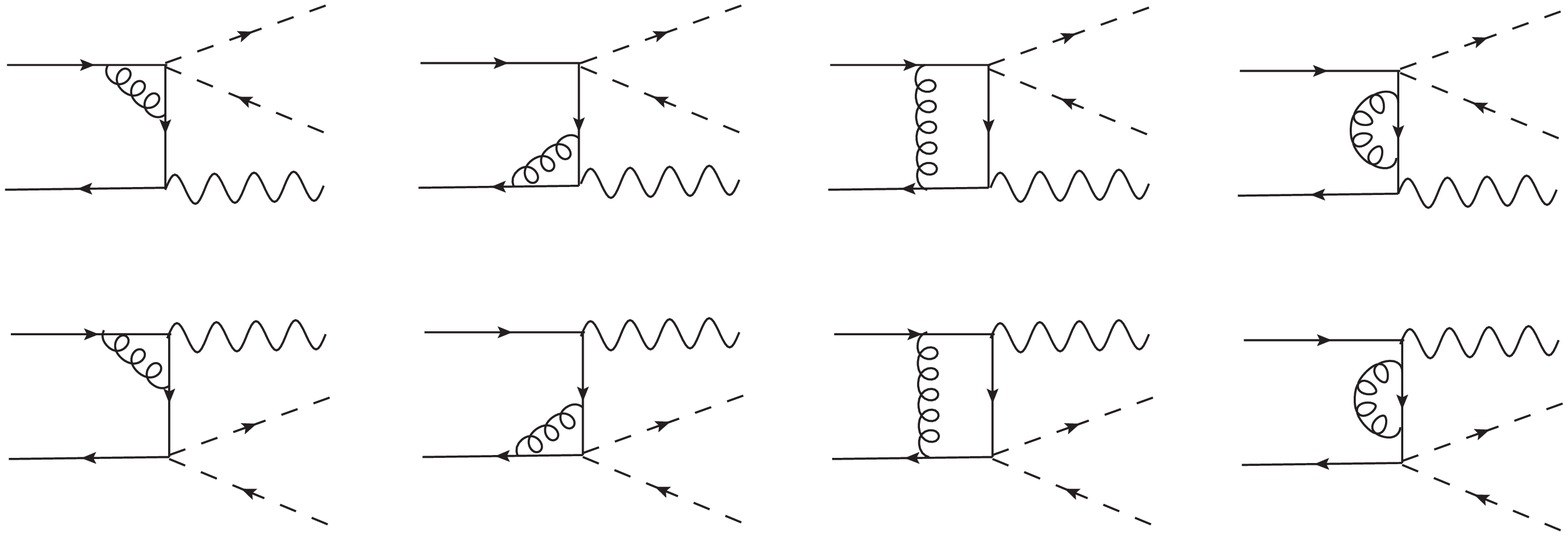}\\
  \caption{Feynman Diagrams for one-loop virtual corrections.}
  \label{fig-virt}
\end{figure}

The virtual gluon corrections to DM and photon associated production consist
of the self-energy, vertex and box diagrams, as shown in Fig.\ref{fig-virt}.  The UV divergences are canceled
between the loop diagrams and counterterms which are determined in on-shell
renormalization scheme for external fields and in $\overline{\rm MS}$ scheme
for coupling constants. The final virtual gluon corrections to the partonic cross section are
\begin{equation}\label{eq-VirtPartonic}
    \hat{\sigma}_v=\frac{1}{2s_{12}}\int d \Gamma_3 2\overline{Re(\mathcal{M}_B^{*}\mathcal{M}_v)},
\end{equation}
in which
\begin{eqnarray}
  \mathcal{M}_{v} &=& \frac{\alpha_s }{4\pi}C_{\epsilon}\left[\left(\frac{A_2^v}{\epsilon^2}+
  \frac{A_1^v}{\epsilon^1}+A_0^v\right)\mathcal{M}_B
   + C_F \frac{4 M_3 t_{15}-4 M_2 t_{25}+3 M_1 \left(t_{15}+t_{25}\right)}{t_{15} t_{25}}\mathcal{M}_{DM}\right],\nn\\
\end{eqnarray}
where $C_{\epsilon}=\Gamma(1+\epsilon)[(4\pi\mu_r^2)/s_{12}]^\epsilon$ and
\begin{eqnarray}
  A_2^v &=& -2C_F, \nn\\
  A_1^v &=& -3C_F, \nn\\
  A_0^v &=&C_F\bigg[ 3\ln \frac{\Lambda^2}{s_{12}}+ \ln ^2\left(\frac{s_{12}}{t_{15}}\right)+\ln
   ^2\left(\frac{s_{12}}{t_{25}}\right)+2
   \text{Li}_2\left(-\frac{s_{12}+t_{15}}{t_{25}}\right) \nn\\
   &+&2\text{Li}_2\left(-\frac{s_{12}+t_{25}}{t_{15}}\right)
   +4\text{Li}_2\left(-\frac{t_{15}+t_{25}}{s_{12}}\right)+2 \pi ^2 \bigg].
\end{eqnarray}
We can also write Eq. (\ref{eq-VirtPartonic}) as
\begin{equation}
    d\hat{\sigma}_v=\frac{\alpha_s }{2\pi}C_{\epsilon}\left[\left(\frac{A_2^v}{\epsilon^2}+
  \frac{A_1^v}{\epsilon}+A_0^v\right)d\hat{\sigma}_B
   + d\tilde{\sigma}_v \right],
\end{equation}
with
\begin{equation}
    d\tilde{\sigma}_v=-\frac{1}{2s_{12}} \frac{4\pi\alpha\kappa^2 C_F}{3\Lambda^4} \frac{ 4 s_{12}^2+5 \left(t_{15}+t_{25}\right) s_{12}+3
   \left(t_{15}+t_{25}\right){}^2}{t_{15} t_{25}}|\mathcal{M}_{DM}|^2d \Gamma_3.
\end{equation}
The first term in $A_0^v$ results from the $\overline{\rm MS}$ renormalization
of the four-fermion operator.
The IR divergences remain after renormalization and
have a structure predicted by other methods. They will cancel the IR divergences
coming from real corrections.

\begin{figure}
  \includegraphics[width=0.6\linewidth]{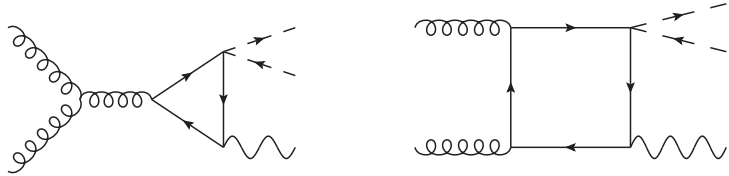}\\
  \caption{Sample Feynman Diagrams for  gluon-gluon initial states contributions.}
  \label{fig-gg}
\end{figure}

A further contribution at hadron colliders arises from the process
\begin{equation}\label{gg}
    g+g \to  \chi+\bar{\chi}+\gamma
\end{equation}
which is illustrated in Fig.\ref{fig-gg}. The effects of these loop-induced gg diagrams are one
 order of $\alpha_s$ higher than that of $q\bar{q}$ diagrams, but they may  be still important
due to the large  gluon PDFs at the LHC. However, the triangle and box diagrams
in Fig.\ref{fig-gg} are both vanishing because of the color structure and charge conjugation symmetry, respectively.

\begin{figure}
  \includegraphics[width=0.6\linewidth]{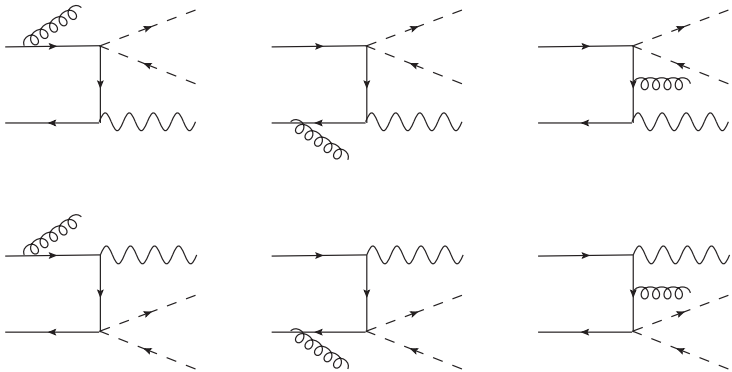}\\
  \caption{Feynman Diagrams for a real gluon emission.}
  \label{fig-real}
\end{figure}

The Feynman diagrams for the real gluon emission process
\begin{equation}\label{eq-GluonEmission}
    q(p_1)+\bar{q}(p_2) \to \chi(p_3)+\bar{\chi}(p_4)+\gamma(p_5)+g(p_6)
\end{equation}
are shown in Fig.\ref{fig-real}. When performing the final states phase space integration, one
encounters the soft and collinear singularities. We use the two cutoff phase space
slicing method to separate the singular regions and perform the integration
analytically in these regions\cite{Harris:2001sx}. Therefore, the real corrections are divided into three
parts, i.e.,
\begin{equation}\label{eq-Real}
    d\hat{\sigma}_r=d\hat{\sigma}_r^S+d\hat{\sigma}_r^{HC}+d\hat{\sigma}_r^{\overline{HC}},
\end{equation}
where $\hat{\sigma}_r^S$ and $\hat{\sigma}_r^{HC}$ denote the contributions from
soft and hard collinear regions, respectively.
The soft regions are defined by the energy of the emitted gluon $E_6\le \delta_s\sqrt{s_{12}}/2$, where $\delta_s$ is the soft cutoff parameter.
The collinear regions are determined according to whether the Mandelstam variables
$t_{i6}=(p_i-p_6)^2$, with $i=1,2$, satisfy the collinear condition $|t_{i6}|<\delta_c s_{12}$, where $\delta_c$ is the collinear cutoff parameter. The hard non-collinear part $\hat{\sigma}_r^{\overline{HC}}$ is finite and
can be computed numerically.

The partonic cross section in soft regions can be factorized as
\begin{equation}\label{eq-Soft}
    d\hat{\sigma}_r^S=(4\pi\alpha_s\mu_r^{2\epsilon})d\hat{\sigma}_B \int dS \Phi_{eik},
\end{equation}
where $dS$ is the integration over the phase space of the soft gluon
\begin{equation}\label{eq-dS}
    dS=\frac{1}{2(2\pi)^{3-2\epsilon}}\int_0^{\delta_s\sqrt{s_{12}}/2}dE_6 E_6^{1-2\epsilon}d\Omega_{2-2\epsilon}
\end{equation}
and the Eikonal factor $\Phi_{eik}$ is the amplitude squared in the soft limit
apart from  $\overline{|\mathcal{M}_B|^2}$, expressed as
\begin{equation}\label{eq-Eik}
\Phi_{eik}=C_F\frac{s_{12}}{t_{16}t_{26}}.
\end{equation}
After integration over the soft gluon phase space, we get
\begin{equation}\label{eq-Soft2}
    d\hat{\sigma}_r^S=d\hat{\sigma}_B\frac{\alpha_s }{2\pi}C_{\epsilon}
    \left(\frac{A_2^S}{\epsilon^2}+\frac{A_1^S}{\epsilon}+A_0^S\right),
\end{equation}
where
\begin{equation}
    A_2^S=2C_F,\qquad A_1^S=-4C_F\ln \delta_s,\qquad A_0^S=C_F\left(4\ln^2 \delta_s-\frac{2\pi^2}{3}\right).
\end{equation}

In hard collinear regions of this process, the momentum of the gluon emitted from initial
partons become collinear to the beam line. In this limit, the four-body matrix elements are
approximated as follows:
\begin{eqnarray}
    \overline{|\mathcal{M}_r|^2} \approx (4\pi\alpha_s\mu_r^2)
    \overline{|\mathcal{M}_B|^2}
    \left[\frac{-2P_{qq}(z,\epsilon)}{z t_{16}}
    +\frac{-2P_{\bar{q}\bar{q}}(z,\epsilon)}{z t_{26}}\right],
\end{eqnarray}
in which, $z$ represents the fraction of initial partons' momentum carried by $q(\bar{q})$.
$P_{ij}(z,\epsilon)$ are the unregulated splitting functions in $n$-dimensions
which can be related to the usual Altarelli-Parisi splitting kernels as
$P_{ij}(z,\epsilon)=P_{ij}(z)+\epsilon P^{'}_{ij}(z)$. In our case,
\begin{eqnarray}\label{eq-AP}
  P_{qq}(z) = C_F\frac{1+z^2}{1-z},\qquad P^{'}_{qq}(z) = -C_F(1-z).
\end{eqnarray}
At the same limit, the four-body phase space can be written as
\begin{equation}
    d \Gamma_4|_{coll}=d \Gamma_3(s^{'}_{12}=z s_{12})
    \frac{(4\pi)^{\epsilon}}{16\pi^2\Gamma(1-\epsilon)}
    dz dt_{16}[-(1-z)t_{16}]^{-\epsilon}.
\end{equation}
Therefore, we obtain
\begin{eqnarray}\label{eq-HChadronic}
  d\sigma_r^{HC} &=& d\hat{\sigma}_B \frac{\alpha_s}{2\pi}C_{\epsilon}
  \left(-\frac{1}{\epsilon}\right)\delta_c^{-\epsilon}[P_{qq}(z,\epsilon)G_{q/p}(x_1/z)
  G_{\bar{q}/p}(x_2) \nn \\
  &+& P_{\bar{q}\bar{q}}(z,\epsilon)G_{\bar{q}/p}(x_1)
  G_{\bar{q}/p}(x_2/z)  + (x_1 \leftrightarrow x_2)]
  \frac{dz}{z}\left( \frac{1-z}{z} \right)^{-\epsilon}dx_1 dx_2.
\end{eqnarray}
To factorize the collinear singularity into the PDFs, we use scale dependent
PDFs in the $\overline{\rm MS}$ convention:
\begin{equation}\label{eq-Redifinepdf}
    G_{b/p}(x,\mu_f)=G_{b/p}(x)+\left(-\frac{1}{\epsilon}\right)
    \left[\frac{\alpha_s}{2\pi}\frac{\Gamma(1-\epsilon)}{\Gamma(1-2\epsilon)}
    \left(\frac{4\pi\mu_r^2}{\mu_f^2}\right)^{\epsilon}\right]\int_x^1\frac{dz}{z}
    P_{ba}(z)G_{a/p}(x/z).
\end{equation}
Now, we replace $G_{q(\bar{q})/p}$ in the LO hadronic cross section (\ref{eq-LOhadronic})
and combine the result with the hard collinear contribution (\ref{eq-HChadronic}). The
resulting $\mathcal{O}(\alpha_s)$ expression for the initial state collinear contribution is
\begin{eqnarray}
  d\sigma^{coll} &=& d\hat{\sigma}_B\frac{\alpha_s}{2\pi}C_{\epsilon}
  \Big\{ \tilde{G}_{q/p}(x_1,\mu_f)G_{\bar{q}/p}(x_2,\mu_f)+
  G_{q/p}(x_1,\mu_f)\tilde{G}_{\bar{q}/p}(x_2,\mu_f) \nn \\
  &+& \sum_{a=q,\bar{q}}\Big[ \frac{A_1^{sc}(a\to a g)}{\epsilon} + A_0^{sc}(a\to a g)\Big]
  G_{q/p}(x_1,\mu_f)G_{\bar{q}/p}(x_2,\mu_f)  \nn \\
  &+& (x_1 \leftrightarrow x_2) \Big\} dx_1 dx_2.
\end{eqnarray}
with
\begin{eqnarray}
  A_1^{sc}(q\to q g) &=& C_F(2\ln \delta_s +3/2 ), \nn \\
  A_0^{sc}(q\to q g) &=& A_1^{sc}(q\to q g)\ln
  \Big( \frac{s_{12}}{\mu_f^2}\Big).
\end{eqnarray}
The $\tilde{G}$ functions are given by
\begin{equation}
    \tilde{G}_{b/p}(x,\mu_f)=\sum_{a}\int_x^{1-\delta_s\delta_{ab}}
    \frac{dy}{y} G_{a/p}(x/y,\mu_f)\tilde{P}_{ba}(y)
\end{equation}
with
\begin{equation}
    \tilde{P}_{ba}(y)=P_{ba}(y)\ln \Big( \delta_c \frac{1-y}{y} \frac{s_{12}}{\mu_f^2} \Big)
    -P^{'}_{ba}(y).
\end{equation}

\begin{figure}
  \includegraphics[width=0.6\linewidth]{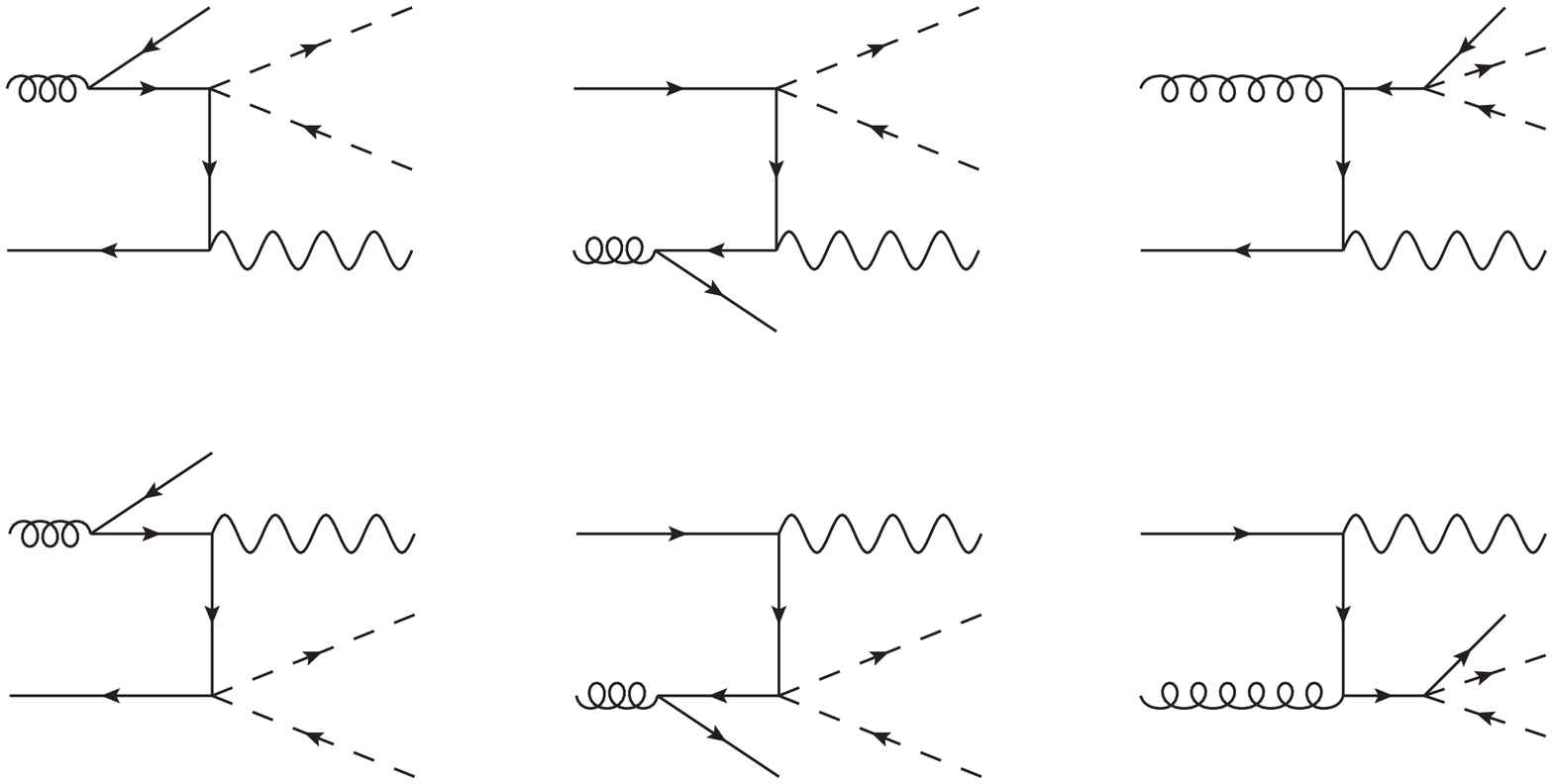}\\
  \caption{Feynman Diagrams for a quark or antiquark  emission.}
  \label{fig-real_qg}
\end{figure}

A complete real correction includes also the (anti)quark emitted processes, as shown in Fig.\ref{fig-real_qg},
such as
\begin{eqnarray}\label{eq-QuarkEmission}
    g(p_1)+\bar{q}(p_2) \to \chi(p_3)+\bar{\chi}(p_4)+\gamma(p_5)+\bar{q}(p_6).
\end{eqnarray}
Their contribution can be obtained from the results of processes in (\ref{eq-GluonEmission}) by
crossing symmetry. Note that, when separating the singular
phase space regions for these processes, one only needs to deal with the collinear
divergences which can be totally absorbed into the redefinition of the PDFs in
(\ref{eq-Redifinepdf}).

Finally, the NLO cross section for the process $pp\to \chi\bar{\chi}\gamma$ is
\begin{eqnarray}
  \sigma^{NLO} &=& \int dx_1 dx_2 \Big\{ \big[ G_{q/p}(x_1,\mu_f)G_{\bar{q}/p}(x_2,\mu_f)
  +(x_1 \leftrightarrow x_2) \big](\hat{\sigma}_B+\hat{\sigma}_v+\hat{\sigma}^S
  +\hat{\sigma}_r^{\overline{HC}}) \Big\}+\sigma^{coll} \nn \\
  &+& \sum_{a=q,\bar{q}}\int dx_1 dx_2 \big[ G_{g/p}(x_1,\mu_f)G_{a/p}(x_2,\mu_f)
  +(x_1 \leftrightarrow x_2) \big]\hat{\sigma_r}^{\overline{C}}
  (ga\to \chi \bar{\chi}\gamma a),
\end{eqnarray}
where $\overline{C}$ in $\hat{\sigma_r}^{\overline{C}}(ga\to \chi \bar{\chi}\gamma a)$ suggests that
the phase space integration is performed in the non-collinear regions. We have checked that
\begin{equation}
    A_2^v+A_2^S=0,\qquad A_1^v+A_1^S+ 2A_1^{sc}(q\to q g)=0.
\end{equation}
Therefore there are no singularities left now.

\subsection{ Numerical results }
In this subsection, we give the numerical results for the cross sections
for DM and photon associated production at the LHC. In numerical calculation, we choose the
CTEQ6L1 (CTEQ6M) PDF sets\cite{Pumplin:2002vw} and the corresponding strong coupling $\alpha_s$ for the LO (NLO) calculations.
The default factorization and renormalization scales, $\mu_f$ and $\mu_r$, are set as $2m$.
We choose the input parameters $(m,\Lambda)=(150{\rm ~GeV},500{\rm ~GeV})$ and $\kappa=1$ unless otherwise specified which are
allowed by the relic abundance constraint.
The kinematic cuts
\begin{eqnarray}
  p_T^{\gamma} &>& 100{\rm ~GeV}, \nn \\
  |\eta^{\gamma}| &<& 2.4,  \nn \\
  p_T^{miss} &>& 100{\rm ~GeV},  
\end{eqnarray}
are applied in our numerical calculation. Here $p_T^{miss}$ is the missing transverse momentum,
defined as
\begin{equation}
    p_T^{miss}\equiv \left\{
                       \begin{array}{ll}
                         p_T^{\gamma}, & \hbox{no jets in the final states,} \\
                         p_T^{\chi\bar{\chi}}, & \hbox{with jets in the final states,}
                       \end{array}
                     \right.
\end{equation}
where $p_T^{\chi\bar{\chi}}$ is the transverse momentum of the system of the DMs.
Jets are defined by the requirements $p_T^{jet} > 20 {\rm ~GeV}$ and $|\eta^{jet}| < 2.5 $.
In order to avoid  QED collinear divergences, we also require the photon to be isolated
by the prescription\cite{Frixione:1998jh}
\begin{equation}
    \sum_{R_{j\gamma}\in R_0} p_T^{jet}< p_T^{\gamma}\Big(\frac{1-\cos R_{j\gamma}}{1-\cos R_0}\Big).
\end{equation}
where $R\equiv \sqrt{\Delta \phi^2+\Delta \eta^2}$ and $R_0=0.4$.

\begin{figure}
  \includegraphics[width=0.6\linewidth]{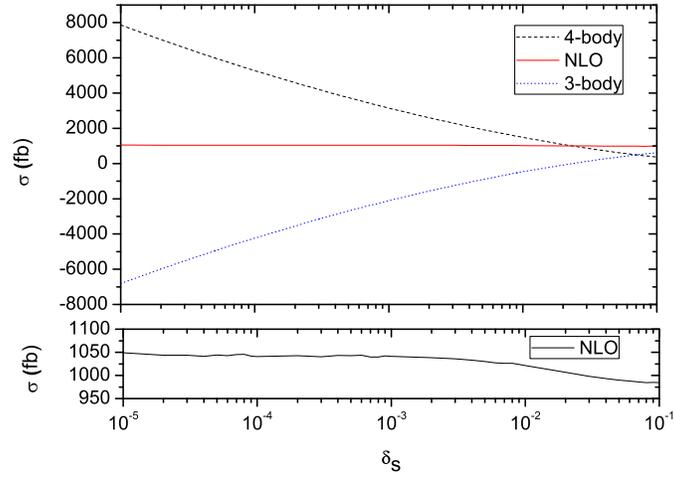}\\
  \caption{Dependence of the NLO cross sections for the DM and photon
  associated production at the LHC on the soft cutoff $\delta_s$ with
  $\delta_c=\delta_s/50$.}
  \label{fig-Cutoff}
\end{figure}

In Fig.\ref{fig-Cutoff} we show the cutoff parameter dependence of the NLO cross sections.
 The contribution of three-body final states
includes the Born cross section, one-loop virtual corrections, soft and collinear limits of the
cross section of four-body final states. The contribution of four-body final states consists in the cross section
of four-body final states with the singular regions of the phase space sliced.
The change of the NLO result is very slow, especially in the region $\delta_s < 10^{-3}$, which
indicates that it is reasonable to use the two cutoff phase space slicing method.

\begin{figure}
  \includegraphics[width=0.48\linewidth]{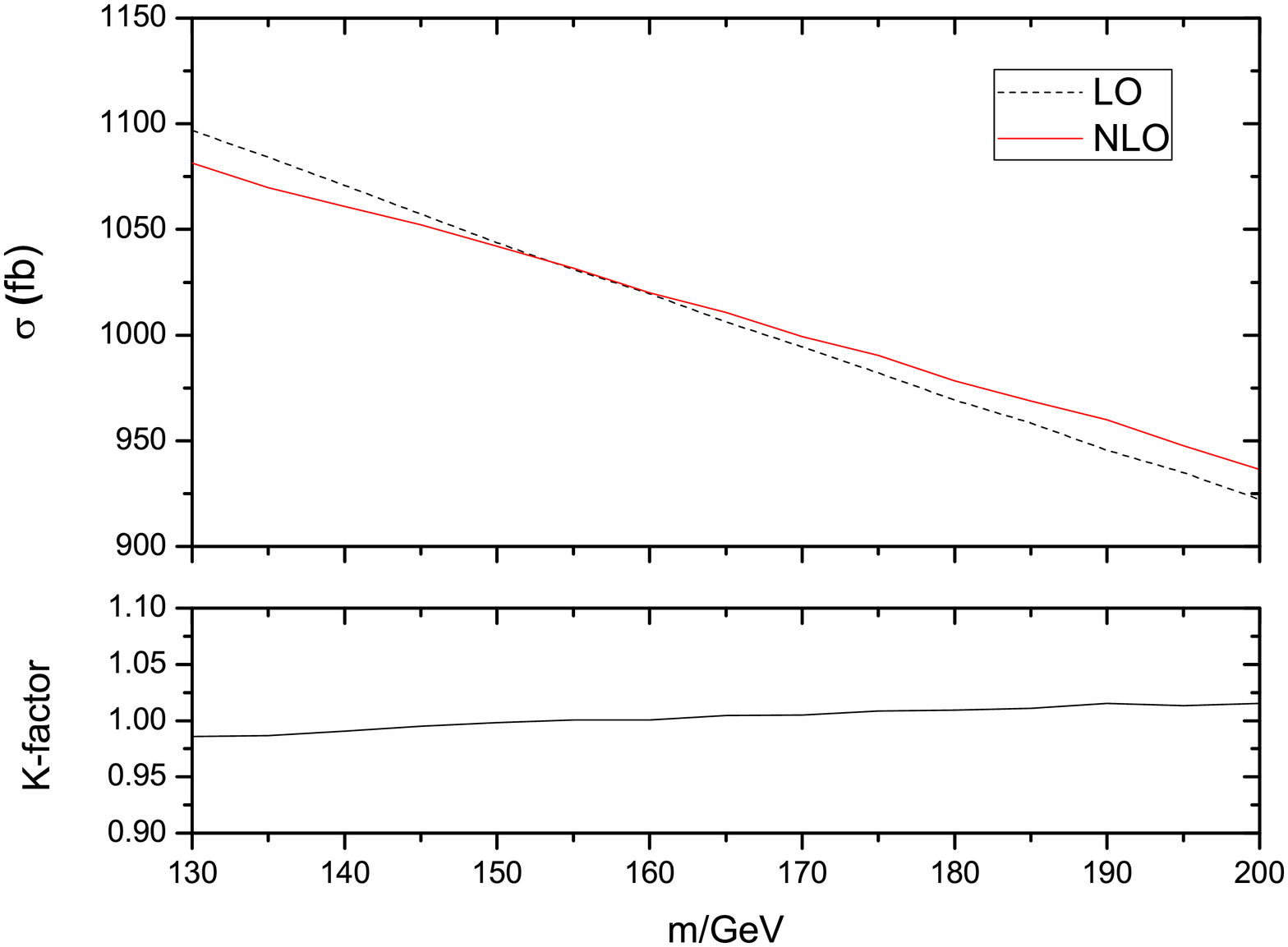}
  \includegraphics[width=0.48\linewidth]{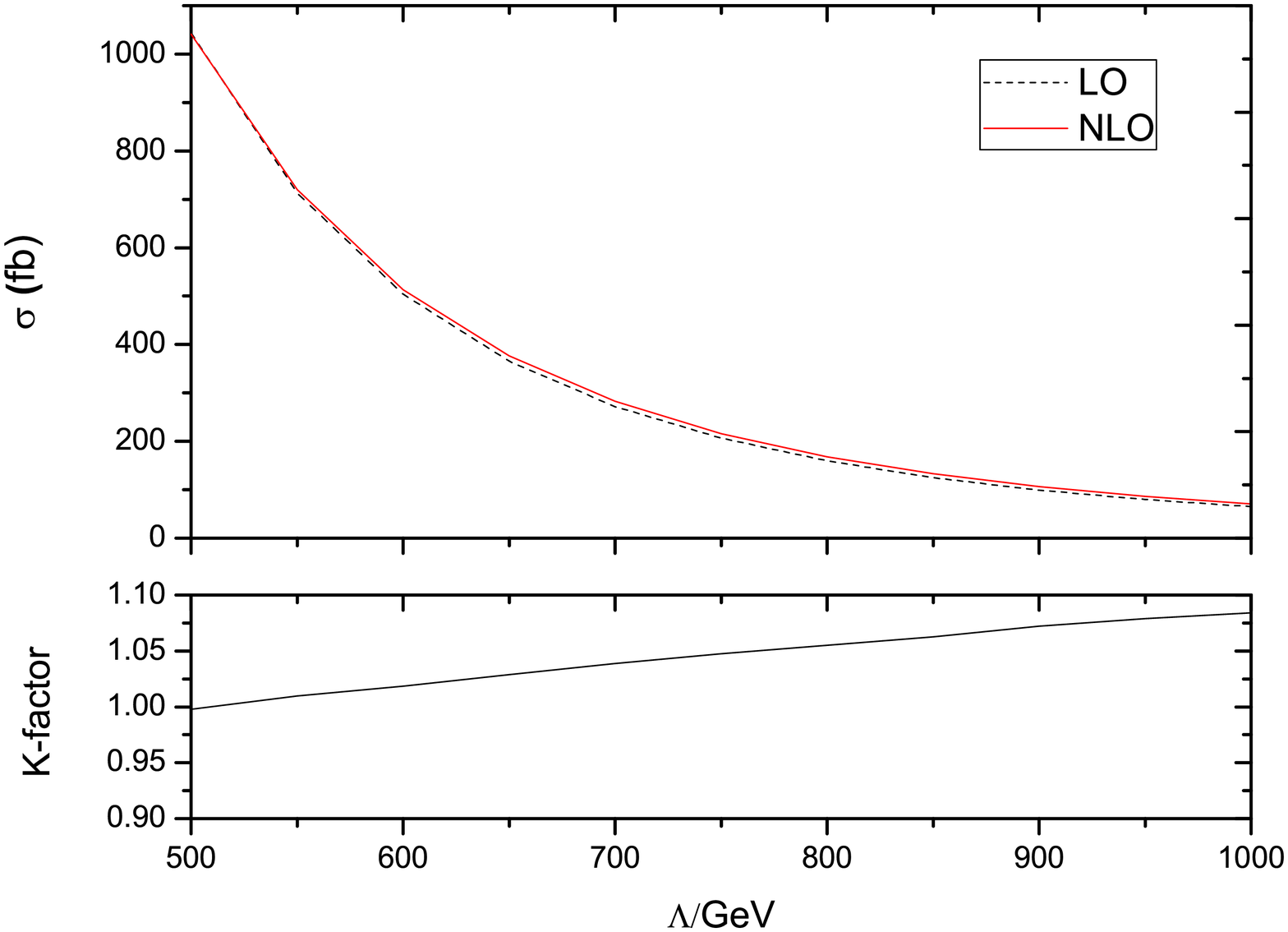}\\
  \caption{Dependence of the LO and NLO cross sections for the DM and photon
  associated production at the LHC on the DM mass and the new physics scale $\Lambda$.
  Also shown is the K-factor.}
  \label{fig-m}
\end{figure}
\begin{figure}
  \includegraphics[width=0.48\linewidth]{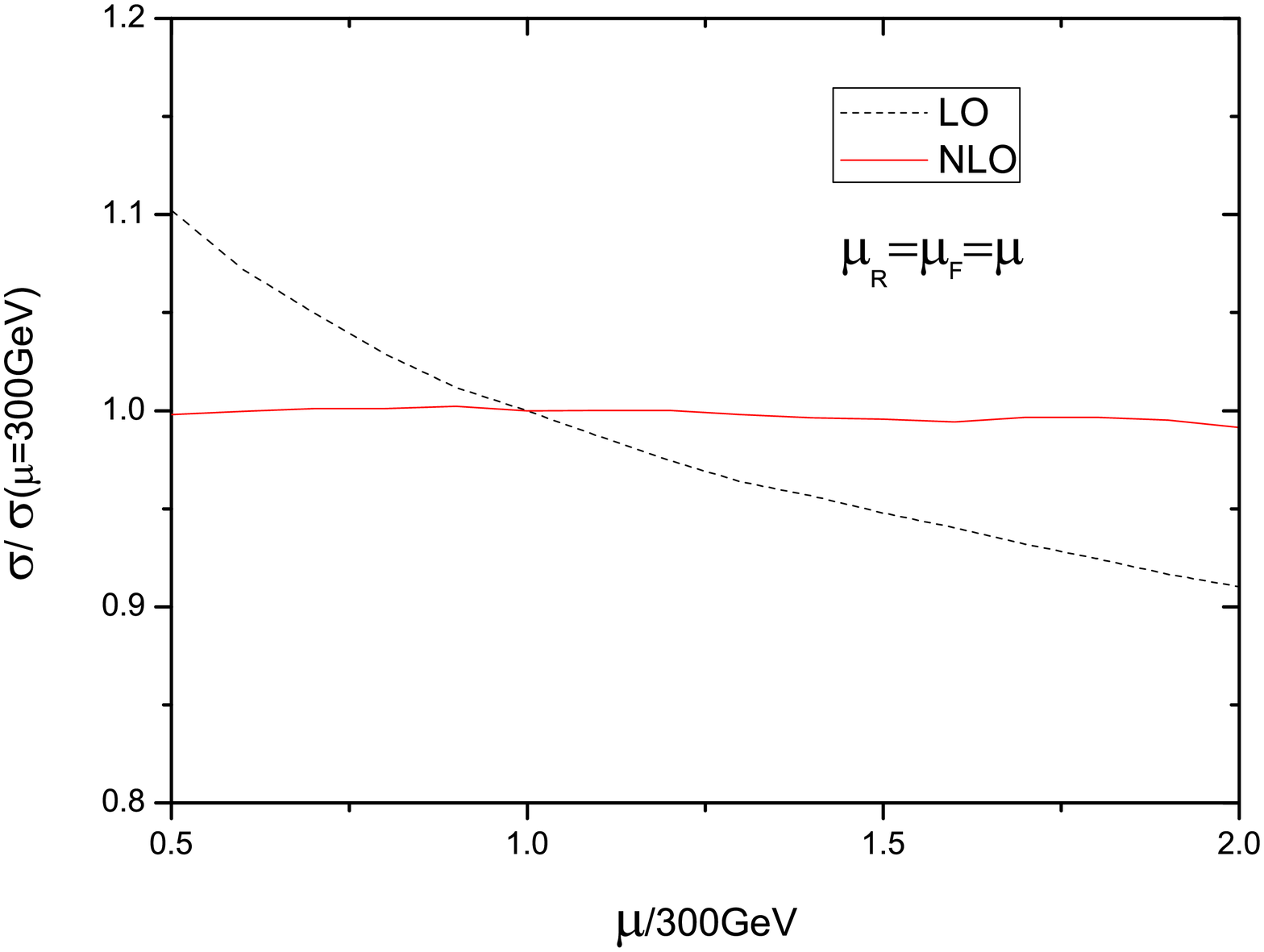}
  \includegraphics[width=0.48\linewidth]{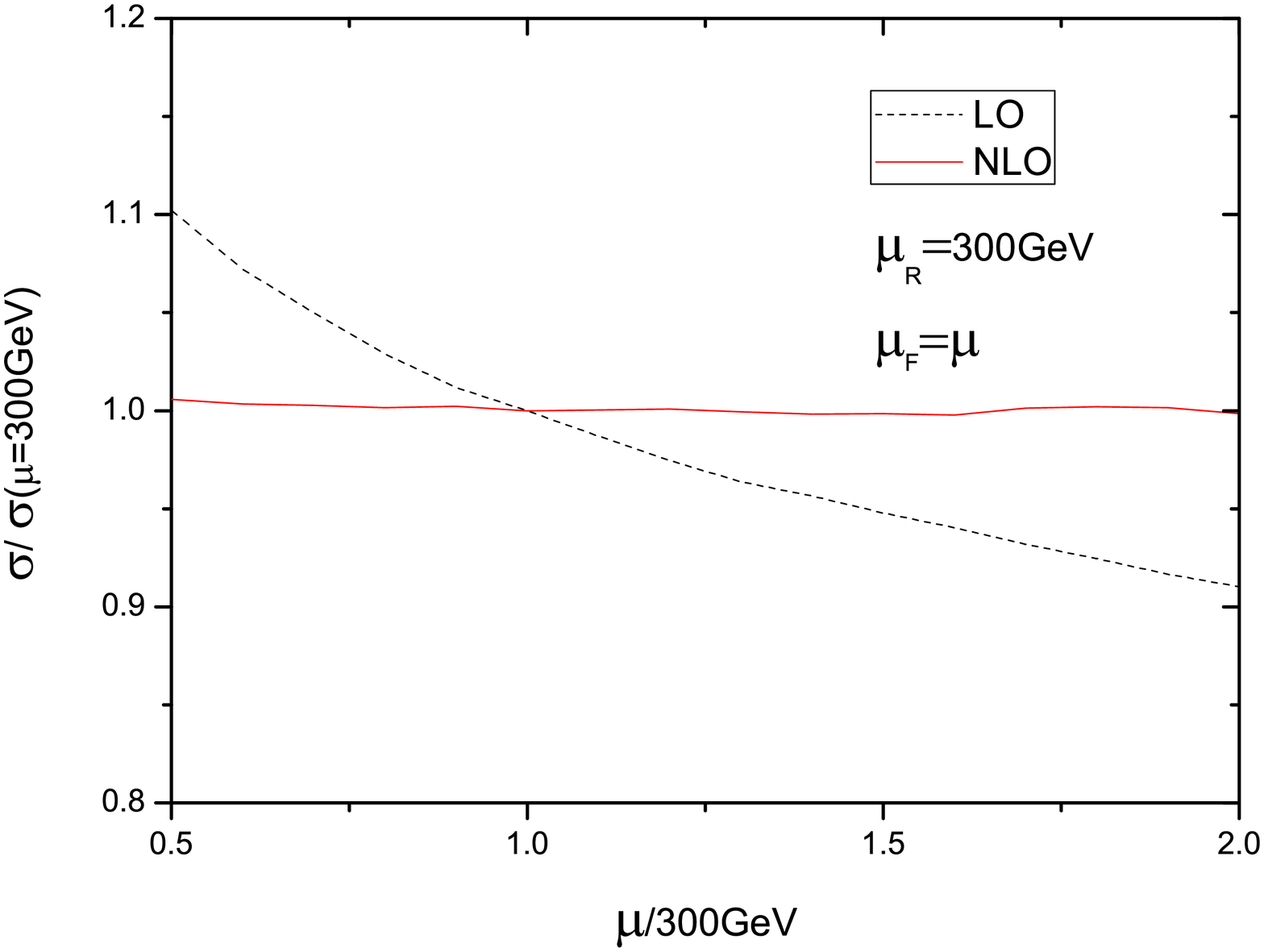}\\
  \caption{Dependence of the LO (NLO) cross sections for the DM and photon
  associated production at the LHC on the factorization scale $\mu_F$ and
  renormalization scale $\mu_R$.}
  \label{fig-FacScale}
\end{figure}

In Fig.\ref{fig-m} we show the $m$ and $\Lambda$ dependence of the LO and NLO cross sections.
The LO (NLO) cross sections decrease from $1097~(1082){\rm ~fb}$ to $922.3~(936.3)~{\rm fb}$
 as $m$ increases from $130{\rm ~GeV}$ to $200{\rm ~GeV}$. The corresponding K-factor,
 defined as the ratio of the NLO cross sections to the LO ones, varies from 0.99
to 1.02. The LO (NLO) cross sections
 decrease from $1044~(1041)~{\rm fb}$ to $65.24~(70.71)~{\rm fb}$
 as $\Lambda$ increases from $500{\rm ~GeV}$ to $1000{\rm ~GeV}$.
 The corresponding K-factor varies from 1.00 to 1.08.
The NLO QCD corrections are modest.
However, the dependence of the NLO cross section on the factorization scale $\mu_F$
and renormalization  scale $\mu_R$
is significantly reduced, as shown in Fig.\ref{fig-FacScale}. This makes the theoretical prediction
much more reliable.

\section{Backgrounds and discovery potential}
\label{sec:background}
The dominant SM backgrounds for this process include the processes $q\bar{q}\to Z(\to \nu\bar{\nu})+\gamma$ and
$q\bar{q}\to Z(\to \nu\bar{\nu})+j$ with the jet misidentified as a photon.
The NLO QCD corrections to these processes are significant. We use
the parton-level Monte Carlo program MCFM\cite{Campbell:2011bn, Baur:1997kz, Ohnemus:1992jn, Giele:1993dj} to estimate these backgrounds at NLO level.
At the Tevatron, the probability $P_{\gamma/j}$ that a jet fakes a photon is almost vanishing
if the transverse momentum of the photon $p_T^{\gamma}$ is larger then $100 {\rm ~GeV}$
because in this situation the hits in the central preradiator chambers are counted and the prompt photon
is distinguished from meson decays\cite{Acosta:2004it}. However, at the LHC, to remain on the safe side,
we set $P_{\gamma/j}=10^{-4}$, as suggested in Ref.\cite{Baur:1992cd}.

\begin{figure}
  \includegraphics[width=0.48\linewidth]{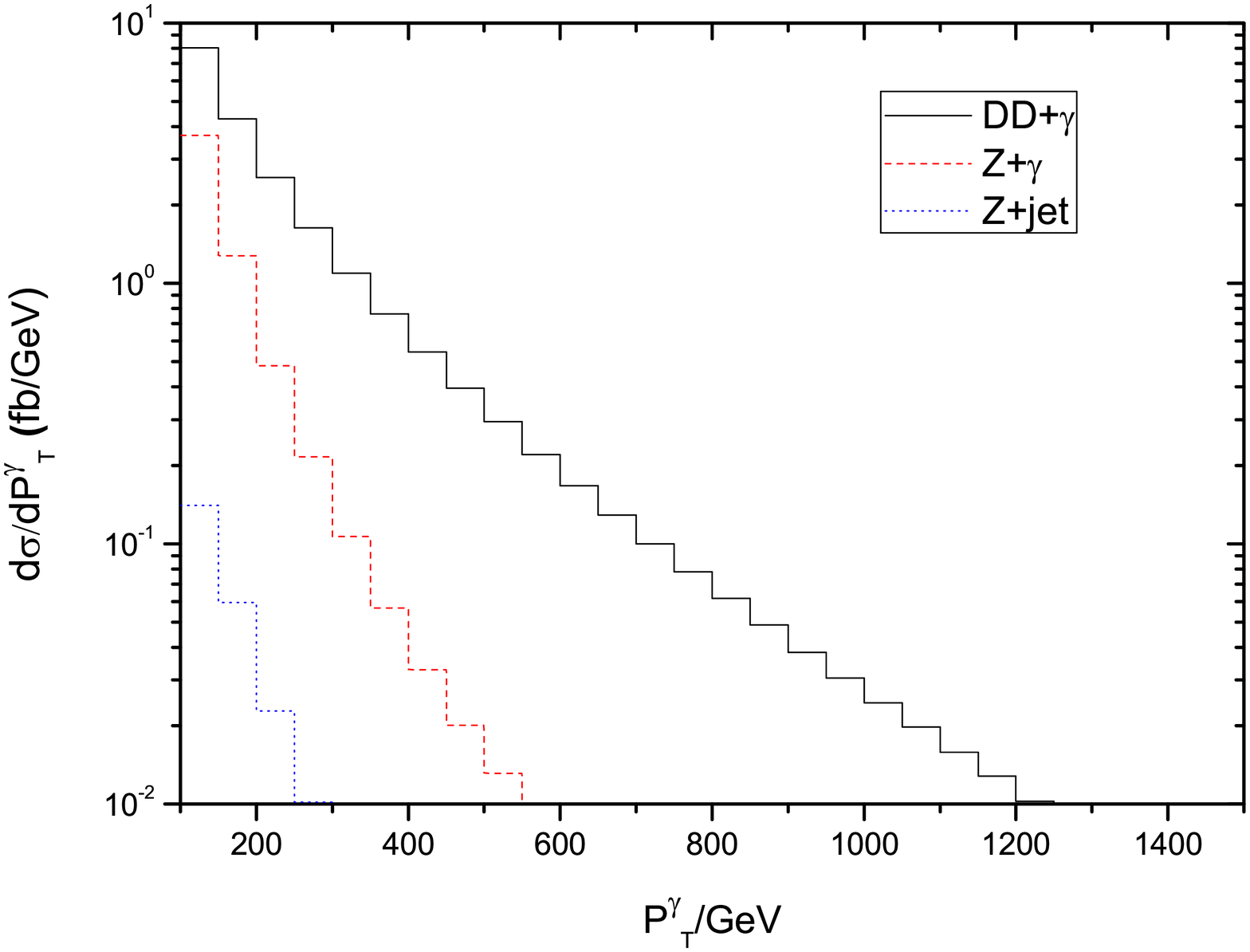}
  \includegraphics[width=0.48\linewidth]{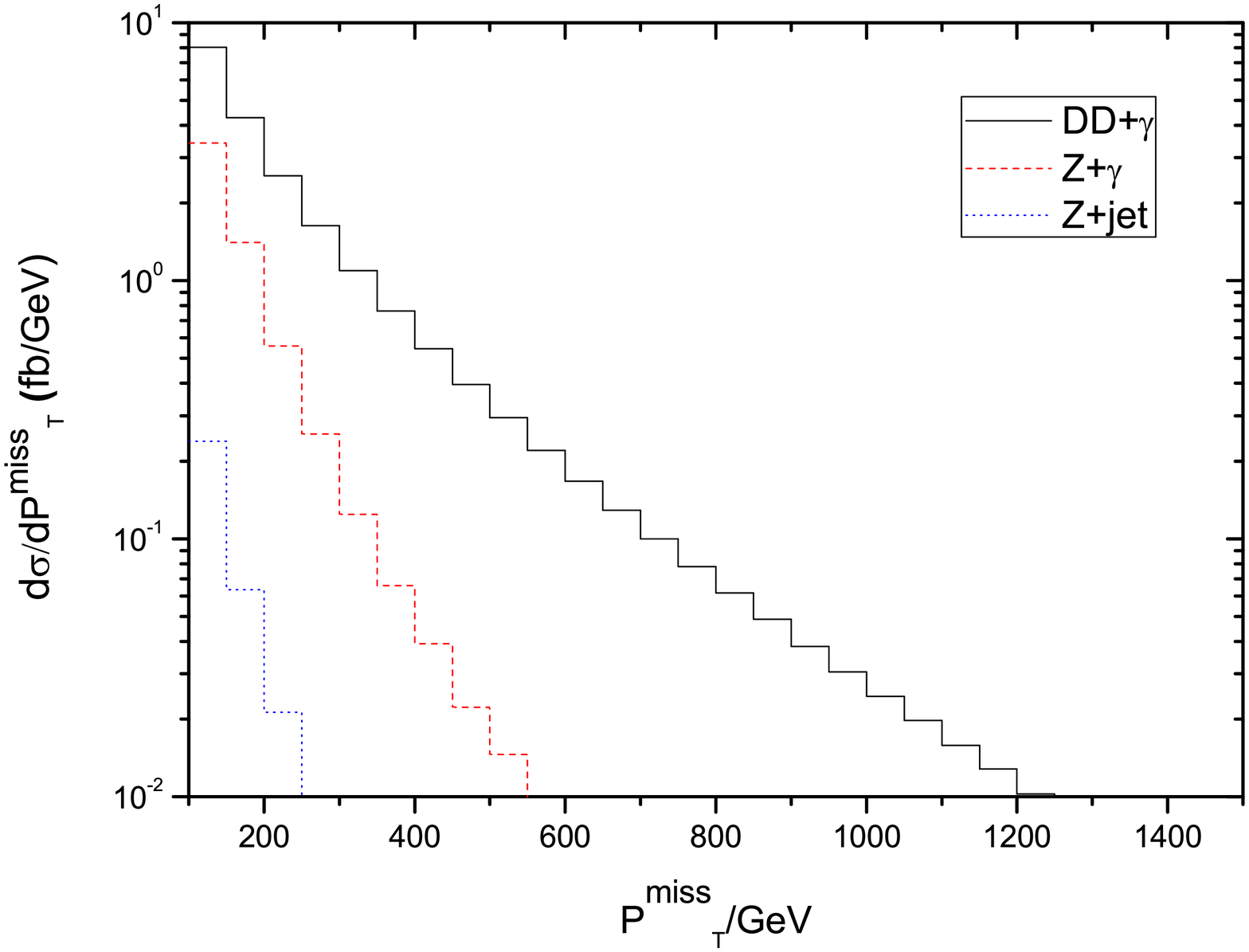}\\
  \caption{Dependence of the differential cross section on $p_T^{\gamma}$ (left) and $p_T^{miss}$ (right).}
  \label{fig-ptgamma}
\end{figure}

Fig.\ref{fig-ptgamma} shows
the differential cross sections of both the signal and backgrounds as a function of $p_T^{\gamma}$ and $p_T^{miss}$.
It can be seen that the $Z\gamma$ production is the main background. The distribution of the
backgrounds decrease faster than that of the signal as the transverse momentum of the photon increases.
Thus, the ratio of signal and background will increase if we set a larger $p_T$ cut.

\begin{figure}
  \includegraphics[width=0.6\linewidth]{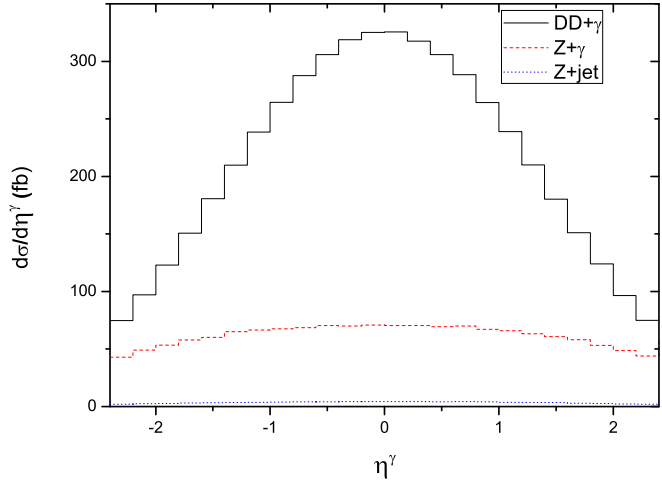}\\
  \caption{Dependence of the differential  cross section on $\eta^{\gamma}$.}
  \label{fig-etagamma}
\end{figure}

Fig.\ref{fig-etagamma} shows
the differential cross section of both the signal and backgrounds as a function of $\eta^{\gamma}$.
The main background is the $Z\gamma$ production, and it is almost flat in the full range of $\eta^{\gamma}$.
In contrast, the signal concentrates in the central region of $\eta^{\gamma}$.
This is a result of the scalar nature of the effective operator.
These characteristics may help to select the events in experiments.


\begin{figure}
  \includegraphics[width=0.48\linewidth]{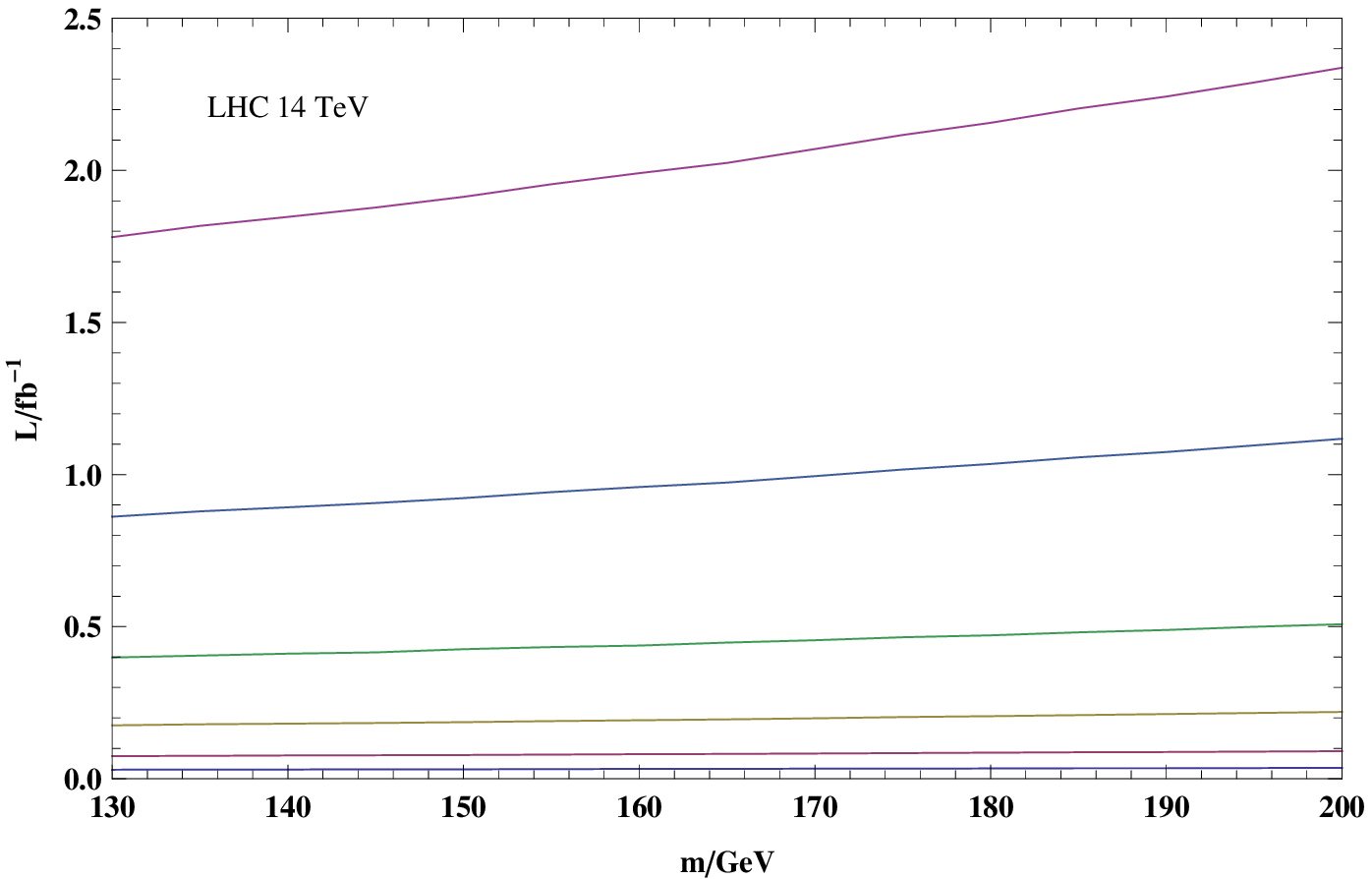}
  \includegraphics[width=0.48\linewidth]{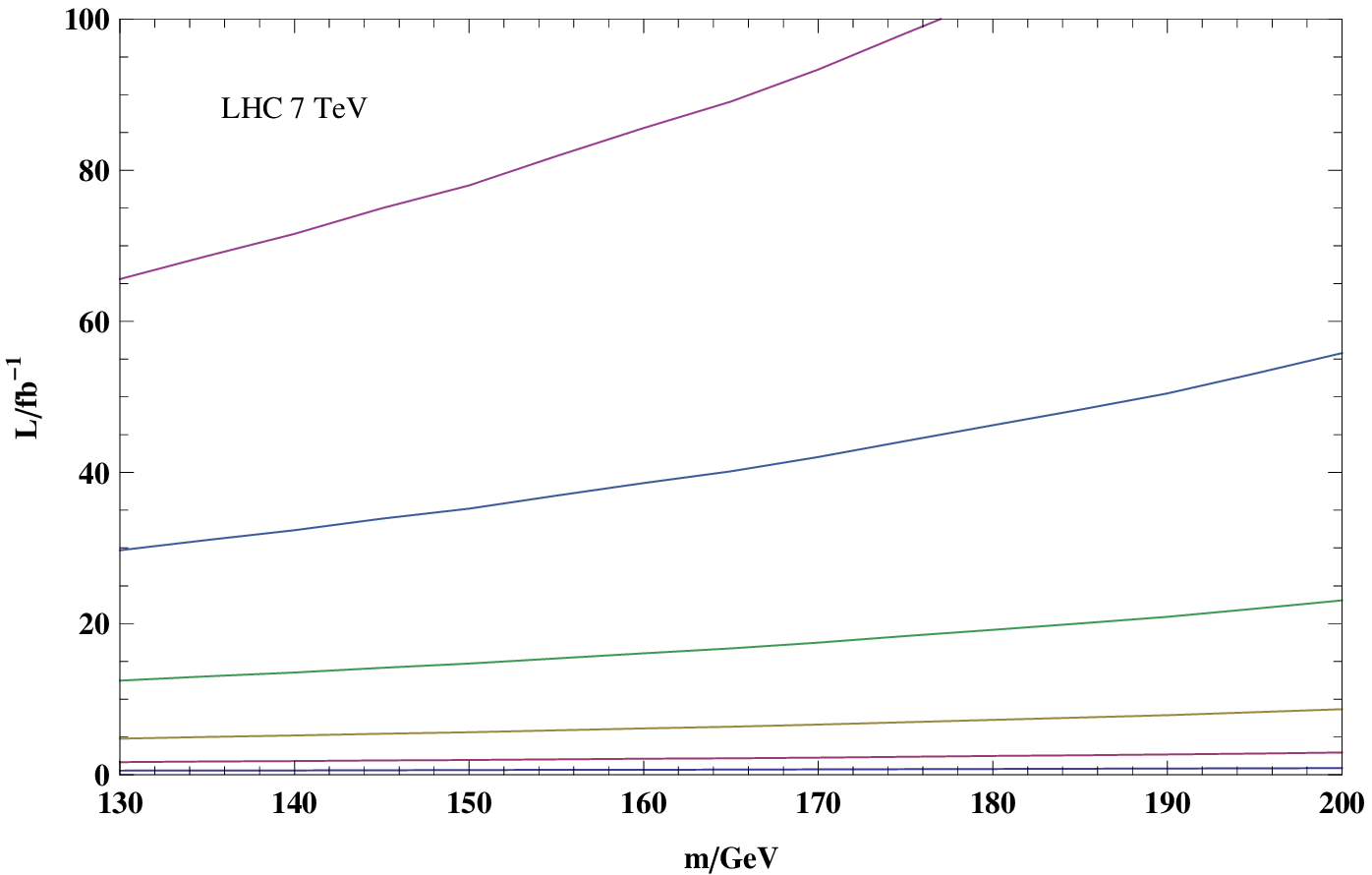}\\
  \caption{The integrated luminosity needed at the LHC with $\sqrt{S}={\rm 14~TeV}$ (left)
  and  $\sqrt{S}={\rm 7~TeV}$ (right)  for a $5\sigma$ discovery.
  The six curves from the bottom up correspond to the new physics scales of 500, 600, 700, 800, 900, 1000 GeV.}
  \label{fig-sig14}
\end{figure}

\begin{figure}
  \includegraphics[width=0.7\linewidth]{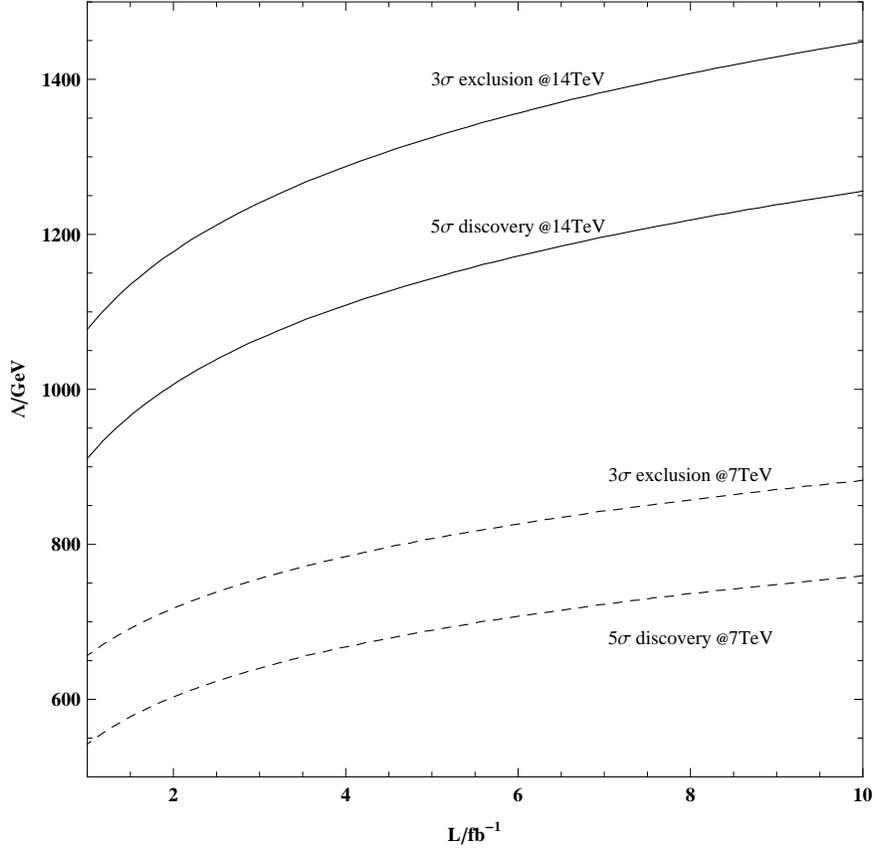}\\
  \caption{The limits of the new physics scale for a $3\sigma$ exclusion and $5\sigma$ discovery at the LHC with $\sqrt{S}={\rm 14~(7)~TeV}$, assuming $m=150$ ~GeV.}
  \label{fig-exclusion}
\end{figure}

Fig.\ref{fig-sig14} shows the integrated luminosity needed at the LHC with
$\sqrt{S}=14 {\rm ~TeV}$ for a $5\sigma$ discovery ($\mathcal{S}/\sqrt{\mathcal{S+B}}=5$) of the signal.
For $\Lambda=500 {\rm
~GeV}$ and $\Lambda=1000{\rm ~GeV}$, the
integrated  luminosities needed are around $0.03~{\rm fb}^{-1}$ and
$2.0~{\rm fb}^{-1}$, respectively. The situation at the LHC with
$\sqrt{S}=7{\rm ~TeV}$ is also shown in Fig.\ref{fig-sig14}.
We find that the LHC may detect this signal once it collects an integrated
luminosity of $1~{\rm fb}^{-1}$, which means that
we may expect the observation of this signal at the early stage of the LHC.
From an experimental point of view, if we discover this signal then
we can set an upper limit for the new physics scale, which has been
illustrated in Fig.\ref{fig-exclusion}.

On the other hand, the LHC may not detect this signal at all. Thus we also present the
exclusion limits of the new physics scale
at the $3\sigma$ ($\mathcal{S}/\sqrt{\mathcal{B}}=3$) level in Fig.\ref{fig-exclusion}.
We can see that the new physics scale is constrained to be larger than $1450~(840)~{\rm GeV}$
if the LHC with  $\sqrt{S}=14~(7){\rm ~TeV}$ does not detect this signal after collecting an integrated
luminosity of $10~{\rm fb}^{-1}$.

\section{Conclusion}\label{sec:conclusion}
We have investigated DM annihilation and signal of DM and a photon
associated production at the LHC induced by a dimension six
effective operator at the NLO QCD level.  We also study the main
backgrounds from SM to this signal, i.e. $Z$ boson and a photon
associated production with invisible decay of $Z$ boson, and $Z$ boson
and a jet production with the jet misidentified as a photon. We find
that the $p_T^{\gamma}$ distributions of the backgrounds decrease
faster than that of the signal  with increasing of the transverse
momentum of the photon. The $\eta^{\gamma}$ distributions of the
backgrounds are almost flat in the full range of $\eta^{\gamma}$.
In contrast, the signal lies mainly in the central region of
$\eta^{\gamma}$. These characteristics may help to select the events
in experiments. We show that in the parameter space allowed by the
relic abundance constraint, which we have calculated at the NLO QCD
level, the LHC  with $\sqrt{S}=7{\rm ~TeV}$ may discovery this
signal at the $5\sigma$ level after collecting an integrated
luminosity of $1~{\rm fb}^{-1}$. On the other hand, if this signal is not
observed at the LHC, we set a lower limit on the new physics scale at the $3\sigma$ level.

\acknowledgments
We would like to thank Qing-Hong Cao for useful discussion.
This work was supported in part by the National Natural
Science Foundation of China, under Grants No.~11021092
and No.~10975004.

\newpage 
\bibliography{DDgamma}

\end{document}